%% file: dataTV.tex
\documentclass[letter]{article}

\usepackage[top=1in,bottom=1in,left=1in,right=1in]{geometry}
\usepackage{times}
\usepackage{graphicx}
\usepackage{url}
\usepackage{helvet}
\usepackage{sectsty}

\allsectionsfont{\sffamily}

\title{DataTV: Streaming Data Videos for Storytelling}

\author{Zhenpeng Zhao, Niklas Elmqvist\thanks{e-mail: elm@umd.edu}\\ %
        \small University of Maryland, College Park%
}

\date{March 2016}

\begin{document}

\maketitle

%% Abstract section.
\abstract{Data videos---motion graphics that incorporate visualizations---have been recognized as an effective way to communicate ideas, but creating such video requires both time and expertise, precluding them from being created and streamed live.
We introduce DataTV, a system for combining multiple media sources in real time.
We validate our work through an expert review involving researchers using the DataTV prototype to create a one-minute data video for their current project.
Results show that the new method facilitates rapid creation and enables users to focus on the narrative rather than mechanics of video production.
}

\begin{figure*}
  \centering
  \includegraphics[width=\textwidth]{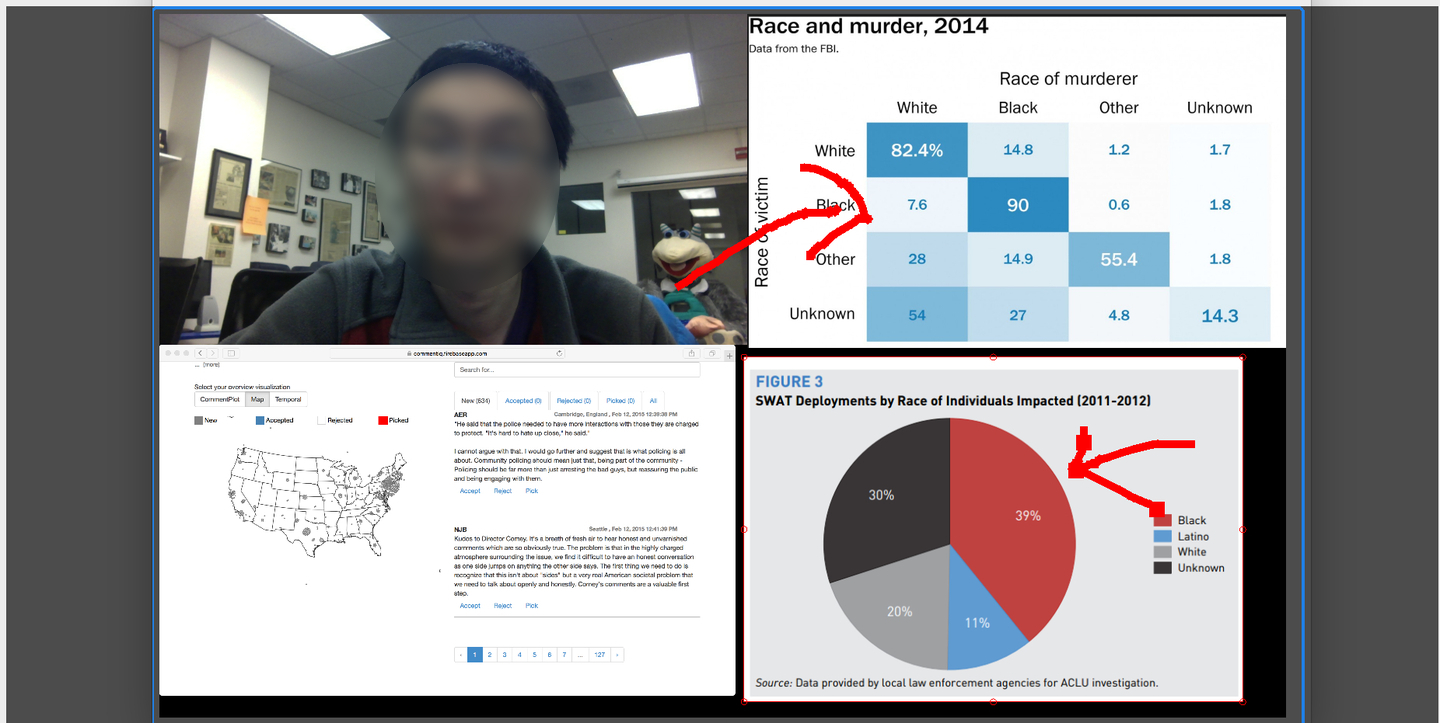}
  \caption{The DataTV prototype tool being used to record a data video on race and murder in the United States in live mode.
  The main viewport is the composition window that shows the current video output being recorded and streamed.}
  \label{fig:teaser}
\end{figure*}

\section{Introduction}
Applying the age-old practice of storytelling to data visualization~\cite{Gershon2001} is one of the most transformative developments in recent history for visualization research~\cite{Kosara2013b}.
As this approach matures, practitioners and researchers alike are adding to the types of storytelling media being used for this purpose~\cite{Segel2010}; beyond the traditional slideshow~\cite{Kosara2013}, recent efforts are now exploring the use of infographics~\cite{Segel2010}, hand-drawn comics~\cite{Bach2016}, and interactive sketching~\cite{Lee2013}.
The use of motion graphics, or so-called \textit{data videos}~\cite{Amini2015}, is particularly interesting as it allows for creating compelling visual narratives using rich multimedia sources, such as live video, graphics, text, sound, and music.
However, authoring effective videos that utilize the medium to its fullest is time-consuming, resource-intensive, and requires a significant amount of skill.
In particular, this precludes such videos from being created and streamed live on the internet.

In this paper, we study how to enable dynamically streaming visual exploration for storytelling.
Such a live ``data TV'' approach will require a streamlined and fully integrated software platform where a single person can prepare, record, and produce the digital stream with a minimum of time investment.
To enable the creation of such live data videos, we propose the \textsc{DataTV} platform, a prototype system for authoring live-streaming data videos using an integrated interface.
The prototype supports three separate modes for (1) production, (2) recording, and (3) editing, in a highly streamlined and optimized workflow that allows a single content creator to control the entire process during live streaming.

To validate the DataTV platform, we asked two visualization experts to create a short data video using DataTV on a specific topic to collect real experiences on using the system for storytelling.
Results from this expert review (participant-generated clips are included in the companion video for this paper) indicate that these researchers were impressed with the relative ease they could create these videos.
Interestingly, our findings suggest that the integrated functionality in the DataTV platform enabled the researchers to focus on storytelling rather than the mechanics of video production.
Their feedback informed further improvements to the platform.

%% ---------------------------------------------------------------------
%% Background
%% ---------------------------------------------------------------------
\input{2-background}

%% ---------------------------------------------------------------------
%% Design
%% ---------------------------------------------------------------------
\input{3-design}

%% ---------------------------------------------------------------------
%% Method
%% ---------------------------------------------------------------------
\input{4-method}

%% ---------------------------------------------------------------------
%% Examples
%% ---------------------------------------------------------------------
\input{5-examples}

%% ---------------------------------------------------------------------
%% Evaluation
%% ---------------------------------------------------------------------
\input{6-evaluation}

%% ---------------------------------------------------------------------
%% Discussion
%% ---------------------------------------------------------------------
\input{7-discussion}

%% ---------------------------------------------------------------------
%% Conclusion
%% ---------------------------------------------------------------------
\input{8-conclusion}

%% ---------------------------------------------------------------------
%% References
%% ---------------------------------------------------------------------

\end{document}

%% file: 2-background.tex
% 2-background.tex
\section{Background}
\label{sec:background}

Motion graphics has recently been proposed as a particularly effective method for data-driven storytelling~\cite{Amini2015}.
Here we review the background on visual communication, visual storytelling, and the state-of-the-art in storytelling for data.

\subsection{Production, Presentation, and Dissemination}

The research agenda for visual analytics called for a focus on the production, presentation, and dissemination of analytics results~\cite{Thomas2005}:

\begin{itemize}
  
\item\textbf{The case for communication:} Vi{\'e}gas and Wattenberg remark upon the proclivity of visualization for communication by virtue of its graphical form, and encourage focusing on so-called \textit{communication-minded visualization}~\cite{Viegas2006} where communication enables collaborative analysis.

\item\textbf{Embedded dissemination:} To reach its full potential, communication capabilities should be integrated into the visualization tools themselves~\cite{Thomas2005}; for example, Tableau now incorporates the story points feature~\cite{Kosara2013}, and most commercial tools supports exporting interactive dashboards to the web.

\item\textbf{Literacy and casual viewers:} Presenting insights from data to the masses requires taking the visualization literacy~\cite{Boy2014} of everyday users into account.
Thus, the notion of ``casual visualization''~\cite{Pousman2007} is important.
  
\end{itemize}

\subsection{Visual Storytelling}

\textit{Storytelling} is the conveyance of a sequence of events involving characters and places---\textit{stories}---and has a history spanning thousands of years~\cite{Schank1995}.
A \textit{visual narrative} is a story told primarily using visual media, such as illustrations, photographs, animations, video, and---now---visualization~\cite{Eisner2008,Sless1981}.

\begin{itemize}
\item\textbf{Richness in visuals:} Visual forms of communication, including icons, illustrations, schematics, photographs, and full-motion video, have long been considered to be one of the most compelling storytelling formats~\cite{Chevalier2016,Sless1981}.

\item\textbf{Video tutorials:} Complex skills are today often learned through step-by-step instructions on the internet~\cite{Grossman2010}.
Several projects have endeavored to increase the richness of such video tutorials: ToolClips~\cite{Grossman2010} provide contextual access to video assistance, Pause-and-Play~\cite{Pongnumkul2011} links tutorials to the real application, MixT~\cite{Chi2012} automatically generates video from demonstrations, and a recent approach makes software tutorial videos interactive~\cite{Nguyen2015}.
  
\item\textbf{New storytelling formats:} Beyond traditional providers (HBO, MSNBC, ABC, etc) as well as the new breed (Amazon Prime, Hulu, and Netflix), open video sharing sites such as YouTube have democratized access to visual storytelling.
With such openness comes entirely new storytelling formats that were previously infeasible, such as vlogs (video blogs)~\cite{Griffith2010}, reaction videos~\cite{Anderson2011}, and video commentaries, including the ever-popular Let's Play videos~\cite{Wadeson2013}.

\item\textbf{Live streaming:} An additional recent development in internet video is the increased focus on live, so-called \textit{streaming}, content, where the media is received and presented to the consumer at the same time it is delivered by the provider.
This development was particularly driven by live gameplay content, where prominent gamers---called \textit{streamers}---broadcast their game screen on services such as Twitch.tv.
While streaming is also used for heavily produced events such as eSports tournaments, where professional gamers compete over thousands of dollars in prize money in games such as \textit{Overwatch}, \textit{League of Legends}, and \textit{Dota 2}, the vast majority of live streams on Twitch and elsewhere are created using specialized \textit{broadcasting software} such as OBS, XSplit, GameShow, etc.
Such software are what inspired Data TV, and are designed for creating high-quality streams using an optimized user interface that can be controlled during live broadcast by a single user.
  
\end{itemize}

\subsection{Data-Driven Storytelling}

Combining communication-driven visualization with storytelling yields \textit{data-driven storytelling}: narrative techniques for telling stories about data~\cite{Segel2010}.

\begin{itemize}

\item\textbf{Storytelling in visualization:} Gershon and Page first proposed using storytelling for visualization~\cite{Gershon2001}, and their work has since been followed up by workshops~\cite{Diakopoulos2011,DiMicco2010}, surveys~\cite{Hullman2011,Segel2010}, and even commercial tools~\cite{Kosara2013}.
  
\item\textbf{Multiple storytelling media:} The purview of data-driven storytelling has quickly grown, from dashboards and slideshow presentations~\cite{Kosara2013} to more esoteric formats such as sketching~\cite{Lee2013}, journeys in time and space, and even comics~\cite{Bach2016,Jin2010}.
In their survey of narrative visualization, Segel and Heer~\cite{Segel2010} identify seven genres---magazine style, annotated charts, partitioned posters, flow charts, comic strips, slideshows, and videos---and also suggest that future storytelling approaches may combine genres.
  
\item\textbf{Data videos:} Amini et al.~\cite{Amini2015} recently identified \textit{data videos} as motion graphics combining both sound and visuals to tell a data story.
Pointing to prominent examples from the New York Times and the Guardian, their work encourages professional storytellers to use visuals to craft their narratives.
Their follow-up Data Clips~\cite{Amini2017} work is an authoring tool for creating data-driven clips incorporating visualizations that can be assembled into longer data videos.
While obviously deeply influential to our work, their treatment focuses on the careful and deliberate production of data videos through ideation, sketching, storyboarding, capture, and editing; live streaming is not covered or even considered.
Furthermore, DataTV provides support for including additional content from the streamer's computer, such as video, sound, music, and existing applications (such as Tableau or Spotfire), merely using the ability to capture the user's full desktop.

\end{itemize}

%% file: 3-design.tex
% 3-design.tex

\section{Design: Streaming Data Video Production}

We claim that there is a need for a multimedia platform for creating such \textit{live data videos} at a pace and scale where they can be streamed and uploaded to an online video sharing service, such as YouTube or Twitch.
Here we describe the major design decisions of the DataTV platform we design to meet this need.

\begin{itemize}
\item[D1]\textbf{Standalone application:} We design our tool to be a standalone desktop application rather than a web-based one.
  \begin{itemize}
  \item Video production and real-time streaming requires high performance processing and significant storage.
  \item No existing streaming software is entirely web-based; in fact, some even employ specialized hardware.
  \item\textit{Alternative:} A web-based tool is platform-agnostic, but real-time video capture is resource-intensive.
  \end{itemize}
  
\item[D2]\textbf{Web integration:} We embed a web browser as a capture source to support web-based visualizations.
  \begin{itemize}
  \item Toolkits such as D3~\cite{Bostock2011} have made the web a unified platform for delivering visualization to the masses.
  \item Modern web browsers are full-featured multimedia platforms supporting a wide range of content, including video, sound, speech, vectors, 3D, etc.
  \item\textit{Alternative:} A web-based tool would trivially support the web, but is impractical due to performance needs.
  \end{itemize}
  
\item[D3]\textbf{Optimized workflow:} The tool supports streaming with a single user acting as talent, engineer, and producer.
  \begin{itemize}
  \item Sustainable workflows for creating streams on a weekly or even daily basis must be time-efficient.
  \item Most streamers on Twitch---even established ones with thousands of subscribers---operate alone.
  \item\textit{Alternative:} Abandoning real-time control would prevent streaming.
  \end{itemize}
  
\item[D4]\textbf{Native media support:} We provide source handlers for many media types, such as video, music, webcams, etc.
  \begin{itemize}
  \item Effective data videos incorporate multiple media types beyond ``just'' the visualizations themselves~\cite{Amini2015}.
  \item Capturing directly from on-screen windows trivially enables native support for all applications.
  \item\textit{Alternative:} Using special software for specific media breaks the workflow, requires expertise, and reduces efficiency, but requires integrating all of the media handlers.
  \end{itemize}
  
\item[D5]\textbf{Simplified video production elements:} We design the tool to provide simplified video production operations using easily accessible actions.
  \begin{itemize}
  \item Typical analysts do not have a background in video editing, much less storytelling using motion graphics.
  \item Providing the building blocks of video production will help creators think in terms of storytelling rather than mundane tool operations.
  \item\textit{Alternative:} Dedicated video editors have a richer set of video production elements, but break the workflow.
  \end{itemize}
  
\end{itemize}

%% file: 4-method.tex
% 4-method.tex
\section{DataTV: A Streaming Data Video Editor}

We present \textsc{DataTV}, a prototype data video streaming utility.
DataTV is a standalone desktop tool for multiple platforms that allows recording multiple video sources from any number of windows on the user's desktop.

\begin{figure}[htb]
  \centering
  \frame{\resizebox{\columnwidth}{!}{\includegraphics{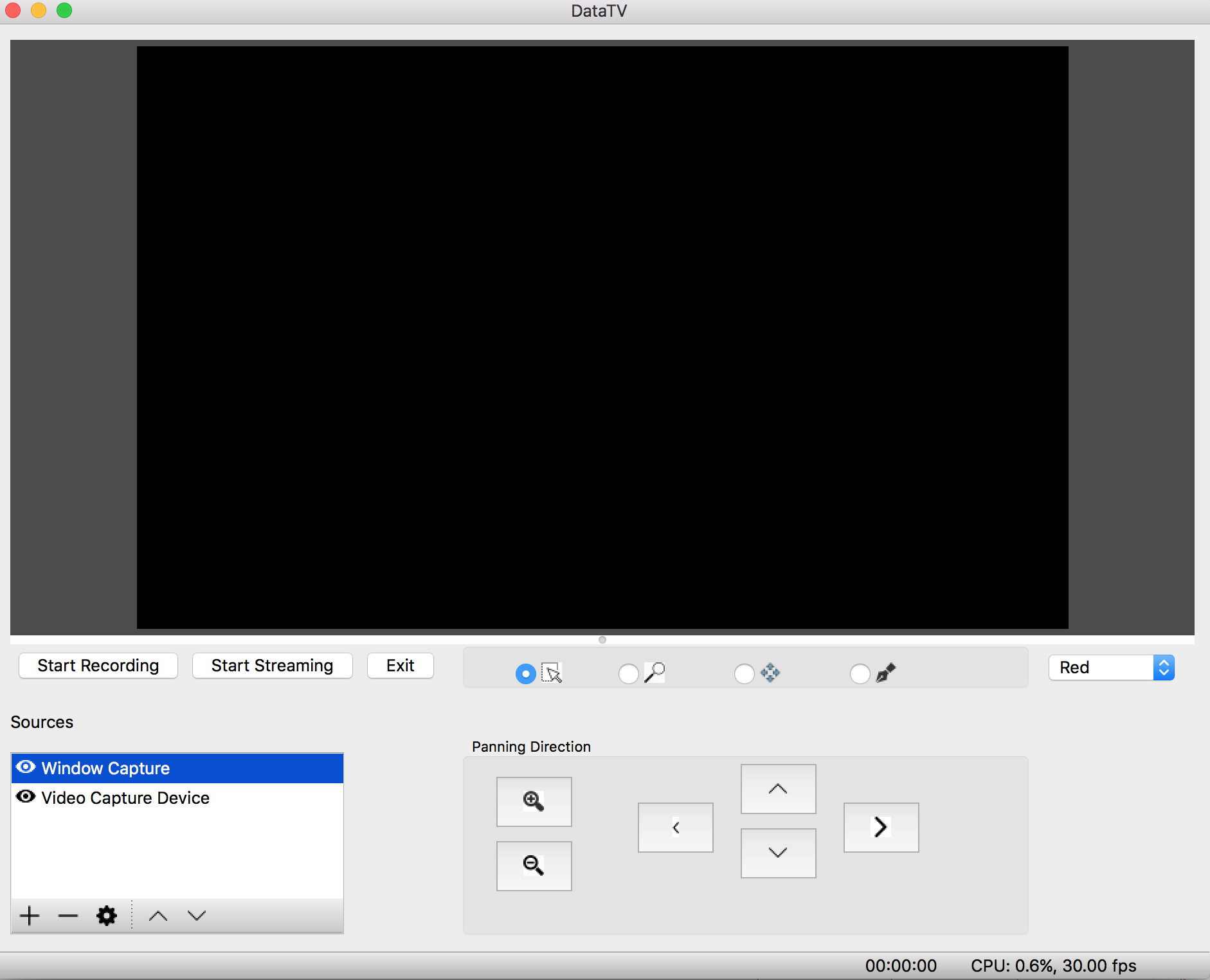}}}
  \caption{Main user interface of the DataTV prototype tool.
  A live mode toolbar allows for panning and zooming sources as well as scribbling directly on top of the output.
  The list panes at the bottom allow for controlling the scenes and sources being displayed.}
  \label{fig:system}
\end{figure}

\subsection{Workflow}

The DataTV tool is modeled along the standard workflow used in game streaming software such as OBS, XSplit, and GameShow, where the streamed output of the tool is managed using the concept of \textit{sources} and \textit{scenes}:

\begin{itemize}

\item\textbf{Source:} Streaming input such as from a window, specific application, audio source, or multimedia content.

\item\textbf{Scene:} Composition of sources on an empty display that can be recorded and streamed as output from the tool.
  
\end{itemize}

DataTV operates in one of two distinct modes: (a) \textit{offline} mode, where the user configures sources and scenes, or (b) \textit{live} mode, where scenes are recorded (and possibly streamed).
Most operations (creating, modifying, and deleting scenes and sources) can be performed in both modes to allow for responding to live events (e.g., adding a new web-based visualization to the stream on a spur-of-the-moment idea), but the most common workflow is as follows:

\begin{enumerate}
\item\textit{Preparation:} The user prepares all scenes and sources for recording in offline mode.
  This involves creating the sources, creating the scenes, and composing the sources for each scene on the output display canvas.
  
\item\textit{Recording:} The user switches to live mode, sets up the stream settings (if enabled), and begins the recording.   
  If the video is streamed live, the user cannot easily switch back to preparation; instead, any changes to scenes or sources must be made while recording.
  If the video is merely being recorded and not streamed, the user can stop recording and go back to preparation, allowing the clips to be edited together at a later stage.

\item\textit{Storytelling:} The user creates the data video in live mode.
  This entails switching between scenes (by selecting the scene to display in the scene manager), managing specific sources (transforming, toggling on and off, annotating, etc), interacting with visualizations and other windows, and potentially narrating and/or capturing webcam video of the user.

\end{enumerate}

\subsection{Source Management}

DataTV maintains a list of currently available sources in the source window (Figure~\ref{fig:system}).
Using this widget, sources can be created, toggled on and off, and deleted in both offline and live modes.
Sources are created by selecting the source type and then associating the source with the appropriate object, such as a specific browser window on the desktop.
Supported sources include the following:

\begin{itemize}
\item\textbf{Window capture:} Video output from a specific window on  the desktop based on the window title, class, or executable.
\item\textbf{Video capture:} Video output from an external device, such as a webcam or video camera.
\item\textbf{Audio capture:} Audio output from a microphone.
\item\textbf{Media objects:} Sources based on static images, video files, and rich text.
\end{itemize}

\subsection{Scene Management}

A \textit{scene} in DataTV is an empty canvas containing sources that form the current output.
Only one scene can be active at a time in DataTV; the active scene is displayed in the main composition window (Figure~\ref{fig:system}) and governs what is recorded and streamed when in live mode.
The user can switch scenes using the scene list.

Managing a scene essentially entails managing the sources involved in the scene.
In offline mode, most scene management operations are ``heavyweight'' in that they require significant setup that is not amenable to live
recording.
Examples of such operations include adding sources to the scene, managing their depth order (governing the drawing order of the sources), and transforming them (translating, scaling, rotating, and cropping).
These operations are achieved by interacting with the sources in the composition window.

In live mode, users should mostly use ``lightweight'' operations that are designed for easy interaction while recording.
This includes toggling the visibility of sources, zooming and panning in a source using mouse dragging and the mouse wheel (such as to zoom in on a particular part of a visualization or window), and scribbling using the annotation feature.

\subsection{Video Annotation}

\begin{figure}[htb]
  \centering
  \frame{\resizebox{\columnwidth}{!}{\includegraphics{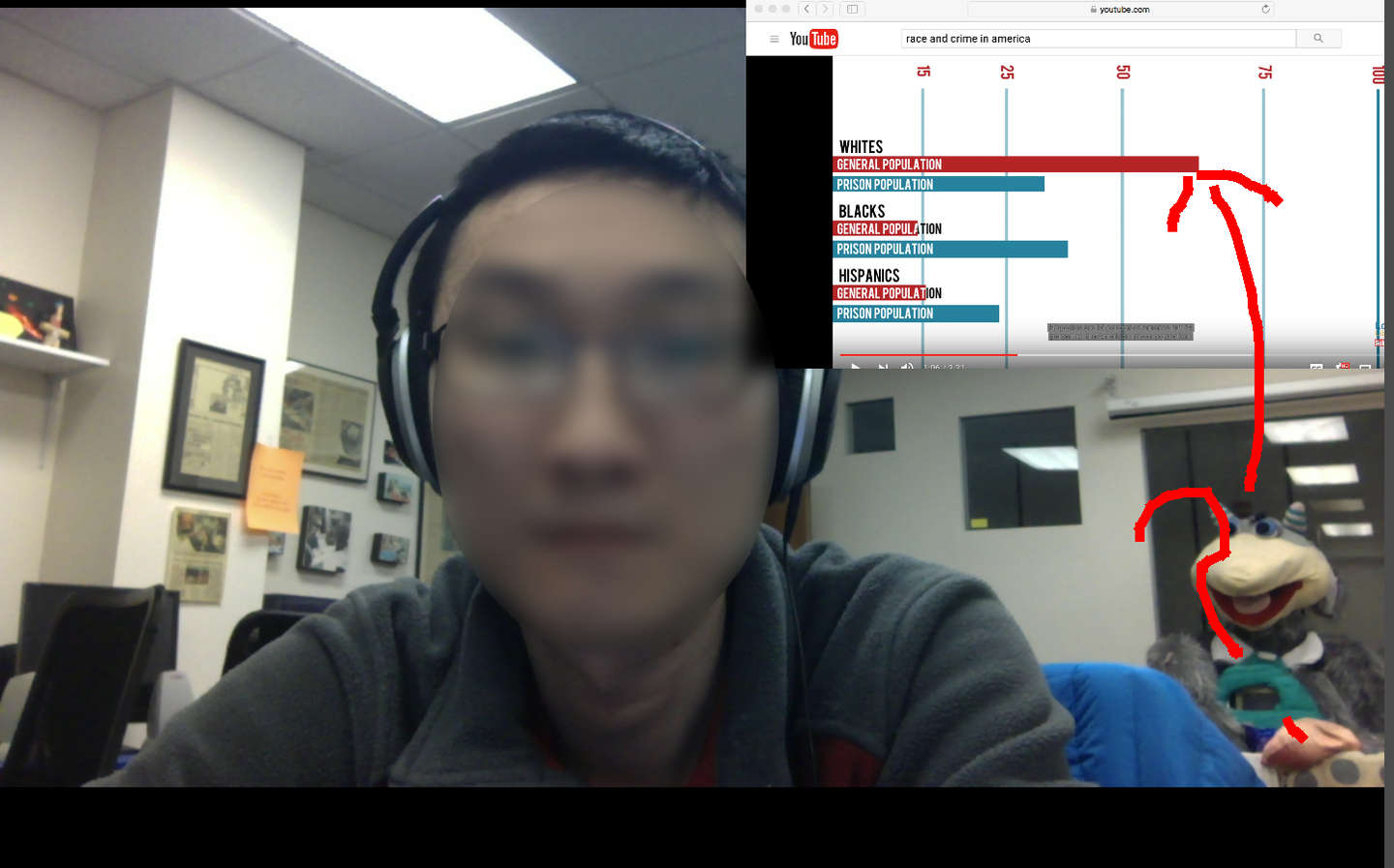}}}
  \caption{DataTV's live mode where users can zoom and pan in a source as well as annotate using pen, eraser, and highlighter.}
  \label{fig:live-mode}
\end{figure}

Sportscasters regularly use annotation to scribble symbols and highlights directly on the video feed, such as to explain specific events in an instant reply of a touchdown or goal.
Presentation software such as Microsoft PowerPoint supports similar ``ink annotations'' where the presenter can draw pen and highlight strokes directly on a slide to illustrate a specific point.
DataTV supports a similar video annotation feature through its live mode interface (Figure~\ref{fig:live-mode}), which provides a pen, highlighter, and eraser tool.
The interface also allows the user to select the drawing color as well as to clear all of the annotation from the output when moving to a new scene.

\subsection{Recording and Live Streaming}

Our DataTV prototype supports recording to standard native video file formats (FLV, MP4, MOV, etc) as well as live streaming output to services such as YouTube and Twitch.tv.
The tool also provides full control over stream settings for both video and audio output.

\subsection{Implementation}

We implemented our DataTV prototype based on OBS (Open Broadcaster Software) Studio, an Open Source live streaming package for multiple platforms.
Our extensions were made in C++ and significantly modifies the workflow of the tool to include a streamlined live mode interface, including scene management, zooming and panning, and live video annotation.

%% file: 5-examples.tex
% 5-examples.tex
\section{DataTV Examples}
\label{sec:examples}

To validate the DataTV prototype and to give a concrete idea of what it is capable of, we present a few examples based on real-life data and existing visualization systems.
The inspiration for these DataTV examples comes from creating and understanding web-based data visualizations and infographics we experience on a daily basis.
Our work here reformulates and reimagines some of the insights from the data video creation process by Amini et al.~\cite{Amini2015, Amini2017}.

% \begin{figure}[htb]
%   \centering
%   \frame{\resizebox{\columnwidth}{!}{\includegraphics{figures/keshif_nobel}}}
%   \caption{Keshif browser for the Nobel Prize Winners dataset.}
%   \label{fig:keshif_nobel}
% \end{figure}

\subsection{Nobel Prize Data Analysis with Keshif}

The first example uses the Keshif~\cite{Yalcin2017} system\footnote{\url{http://keshif.me/}} to explore the ``Nobel Prize Winners'' dataset.
Keshif is a visualization system designed for interacting with multi-dimensional data using a sophisticated faceted browser.
The key interaction in the Keshif system is \textit{linked selection}, which is a generalization of brushing and linking supporting highlighting, filtering, and comparison.

\begin{figure}[htb]
  \centering
  \frame{\resizebox{\columnwidth}{!}{\includegraphics{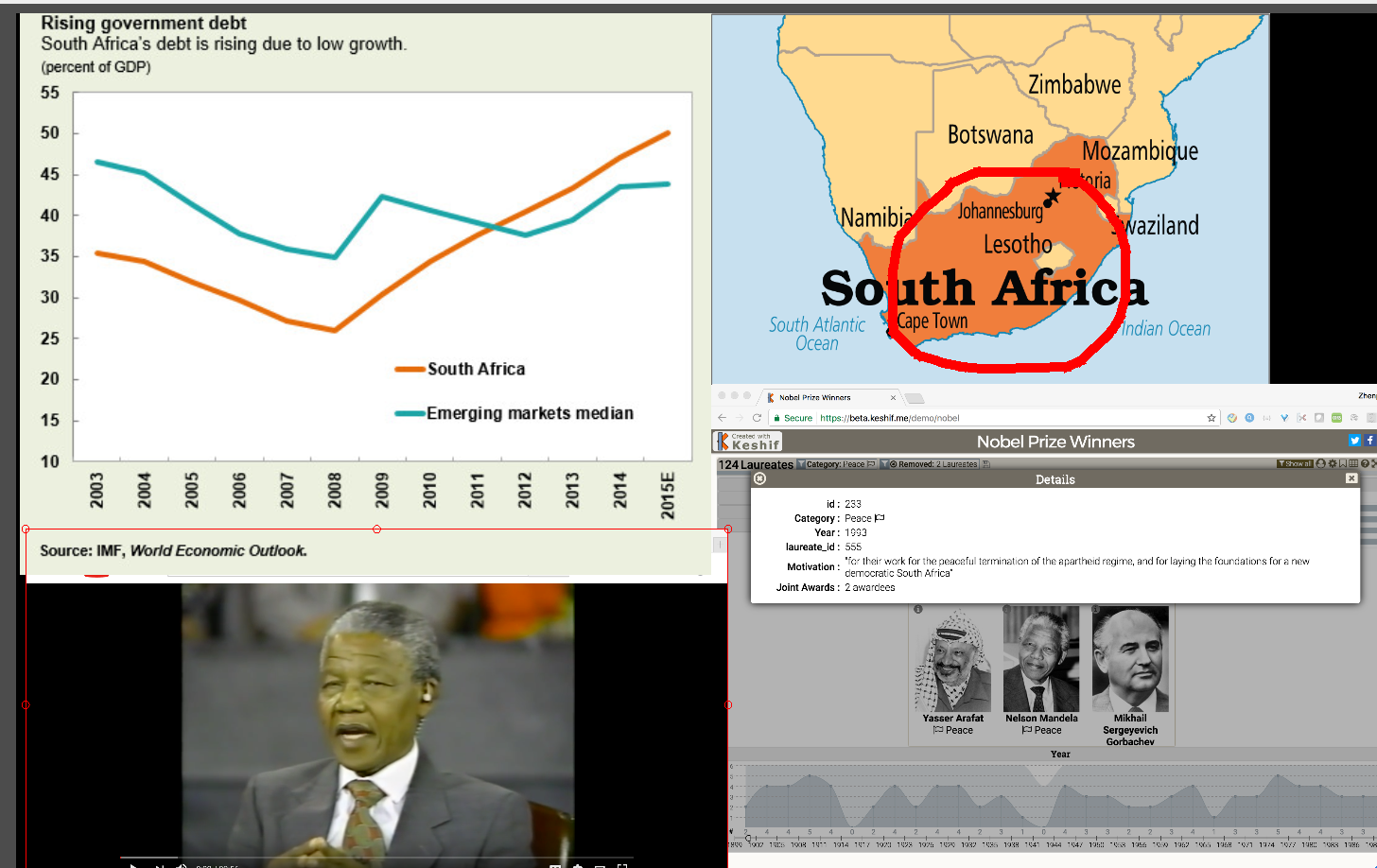}}}
  \caption{A live DataTV session using the Keshif system for the Nobel Prize Winners dataset.
    The user imports an external visualization tool to display the economy of South Africa.}
  \label{fig:keshif_nobel_youtube}
\end{figure}

The Nobel Prize Winners dataset is composed of the basic profiles of all winners, including their pictures, year of winning the prize, nationality, etc.
The overall story of this example is to briefly introduce South Africa, particularly its former leader and Nobel Peace Prize winner, the late Nelson Mandela.
The analyst uses a YouTube video of Mandela (Figure~\ref{fig:keshif_nobel_youtube}) giving a speech as the introduction.
Live video used in this way will make the presentation engaging and draw in the viewer.
It is easily achieved in DataTV using a YouTube source and controlled live by the user; no specific off-line editing is needed.
Then we turn off the video and use a live window with Keshif to lead the story from introducing the categories of Nobel Prize to their age distribution.
Again, the storyteller can do this in real-time simply by interacting with the Keshif visualization in a normal web browser window, potentially  narrating his findings using the webcam and microphone.
The DataTV platform will capture all of these sources, compose them in real-time, and stream them to a remote server.
Compared to static media types, such as infographics, the ability to use an interactive interface enables the analyst to change the topic and approach during the storytelling.
For example, potentially in response to an audience question through the Twitch chat service, the analyst may decide to give a little history of the Nobel Peace Prize, as well as Nelson Mandela's term as president, culminating in him winning the Nobel Prize.
We then conclude the session with a short YouTube video of him giving another speech.

\begin{figure}[htb]
  \centering
  \frame{\resizebox{\columnwidth}{!}{\includegraphics{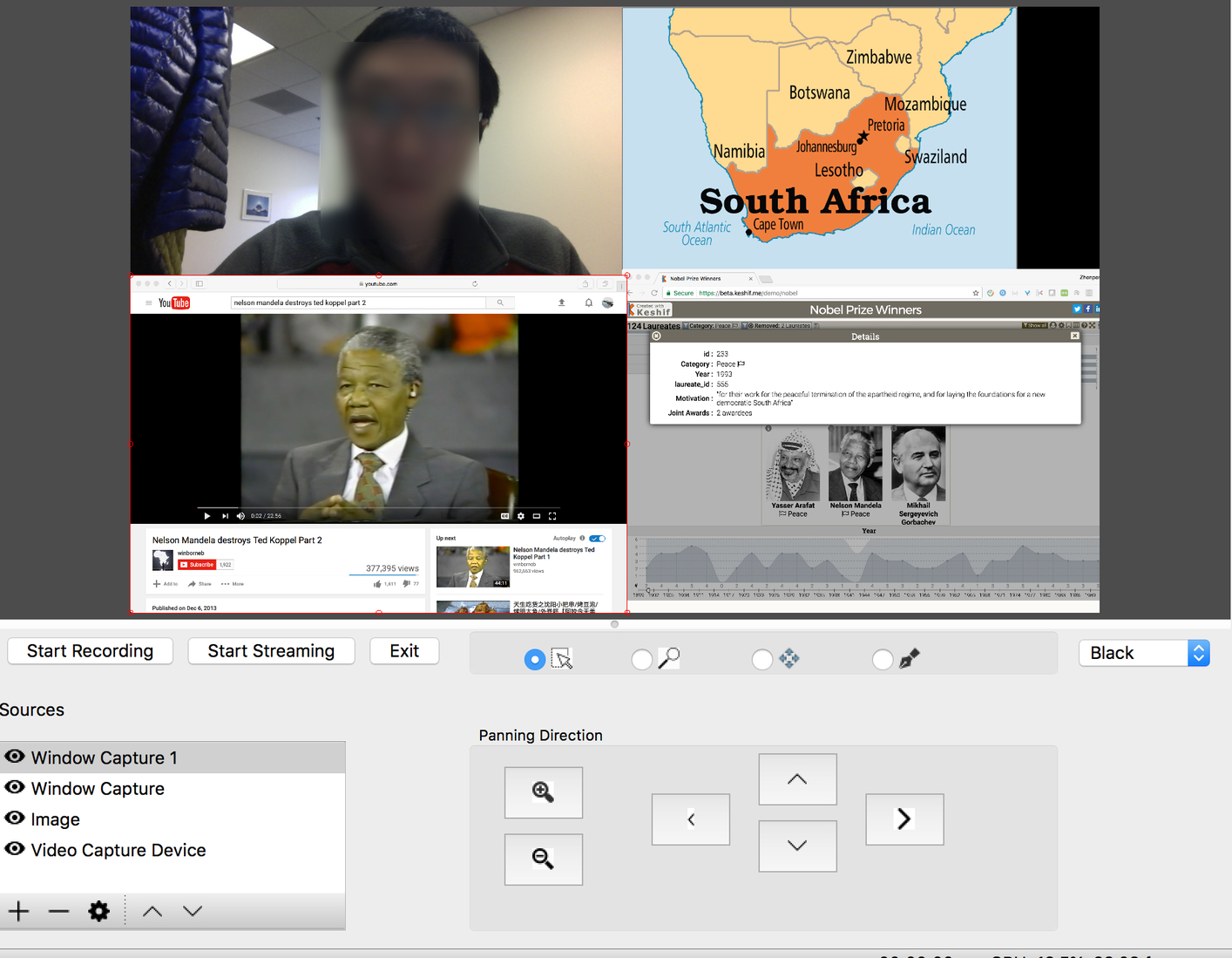}}}
  \caption{Webcam and Keshif visualization being recorded for the Nobel Prize Winner data video.
  	The author is recording a live video ``talking head'' view using a webcam and microphone source as input, which is composed into the final video output.}
  \label{fig:keshif_nobel_interactive}
\end{figure}

% Figure~\ref{fig:keshif_nobel} shows the original interface of the Keshif system, which clearly requires a significant amount of screen space to use properly.
% Since our DataTV prototype supports real-time controls for zooming and scaling, the presenter can easily adapt the size of the browser in the composite video output, even zooming in to show specific features (Figure~\ref{fig:keshif_nobel_interactive}).

\begin{figure}[htb]
  \centering
  \frame{\resizebox{\columnwidth}{!}{\includegraphics{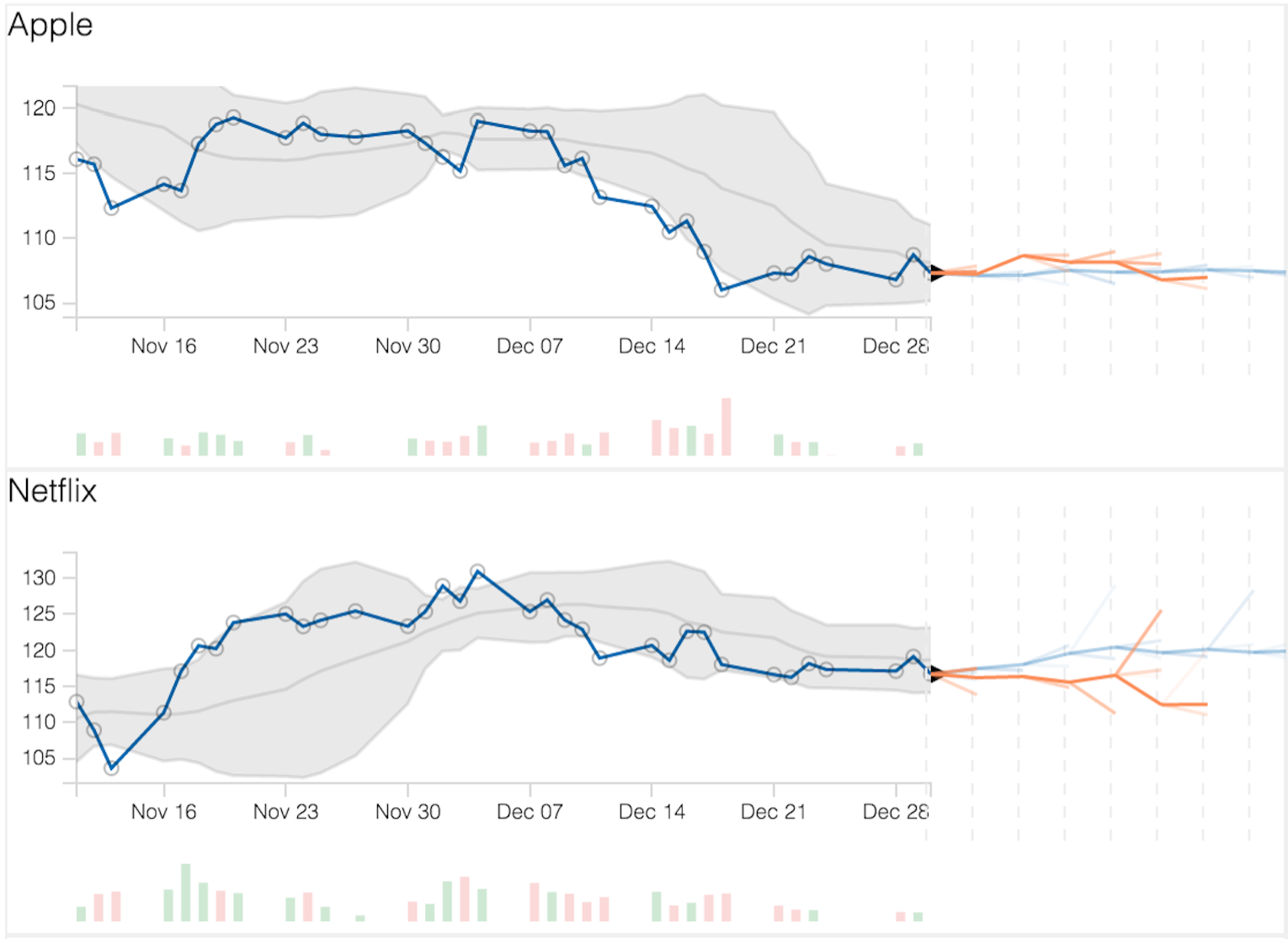}}}
  \caption[]{TimeFork~\cite{Badam2016} prediction space for stock market data.}
  \label{fig:timefork-predition}
\end{figure}

\subsection{Stock Market Data Analysis with TimeFork}

For this example, we use TimeFork~\cite{Badam2016}, an interactive visual prediction technique to support users exploring the future of time-series data. 
The TimeFork implementation allows an analyst to explore a multitude of potential futures for specific stocks by initiating a dialogue between the analyst and the user.
Here we use TimeFork to create a narrative for tech market stocks (here Apple and Netflix).

Our user starts off the session with an introductory scene involving a YouTube video featuring Warren Buffett talking about the current stock market.
Then the user switches to a scene incorporating the TimeFork tool, using it to predict the current trend of hot tech stocks Apple and Netflix by simply interacting with the tool in a web browser.
The user can even switch to a window of a desktop visualization tool such as Tableau or Spotfire and include that into the video if desired.
Meanwhile, the user is narrating the interaction and the findings using a live recording of himself using the computer's webcam and its built-in microphone.  
All of these media sources are composed, recorded, and streamed in the DataTV tool in real-time using the active scene specification.
The main narrative of the data video would describe a scenario for stock market trends, similar to what you may hear on financial news.
The user closes the video with a PowerPoint information slide that summarizes the main trends.

\begin{figure}[htb]
  \centering
  \frame{\resizebox{\columnwidth}{!}{\includegraphics{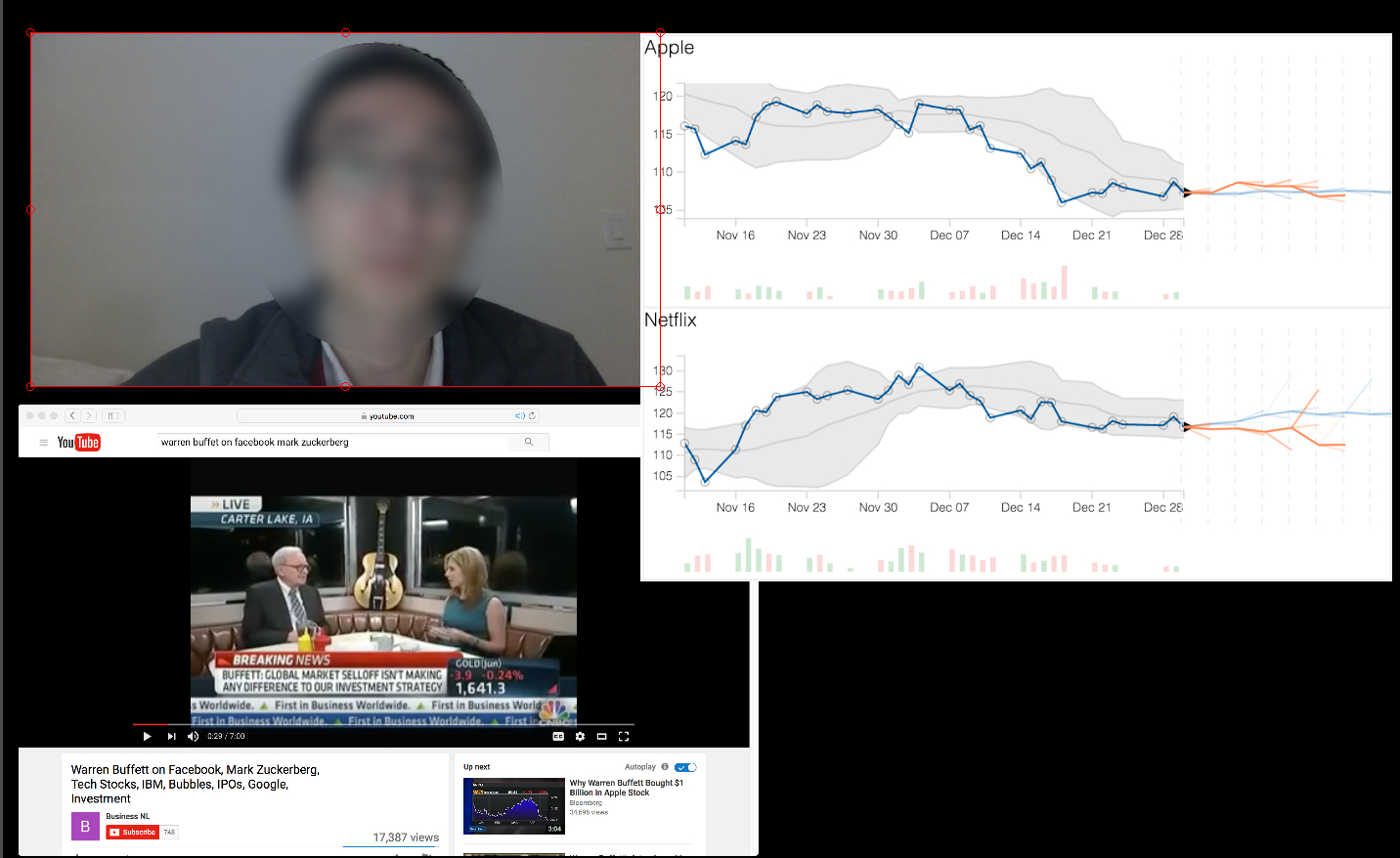}}}
  \caption[]{DataTV recording session for a stock market prediction story involving a YouTube clip of Warren Buffet talking about the current state of the stock market as well as the TimeFork visualization tool~\cite{Badam2016} for predicting the stock price of Apple and Netflix.}
  \label{fig:timefork_youtube}
\end{figure}

\subsection{NY Times Comment Data Analysis with CommentIQ}

The CommentIQ~\cite{Park2016} system is designed to help community moderators manage large amount of comments associated with online articles by automatically ranking them based on criteria such as relevance, readability, personal experience, and length.

\begin{figure}[htb]
  \centering
  \frame{\resizebox{\columnwidth}{!}{\includegraphics{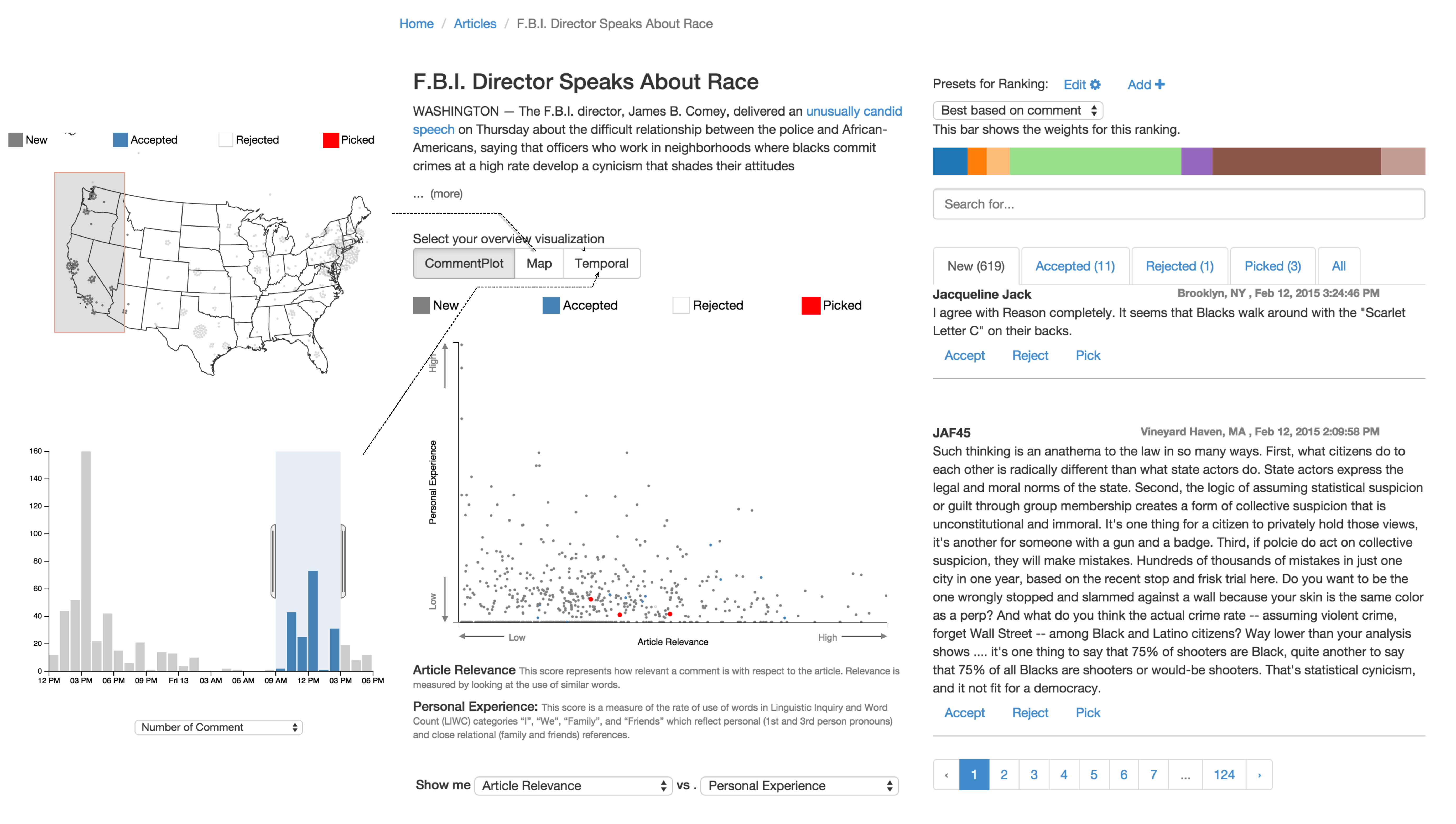}}}
  \caption{Interface of CommentIQ system supporting multidimensional analysis for online article comments.}
  \label{fig:presenter-slide}
\end{figure}

In this example, the user wants to author a streaming data video about the community response on an article from The New York Times\footnote{\url{http://www.nytimes.com/}} titled ``City Reacts: State of Emergency'' during the 2015 racial unrest in Ferguson, Missouri following the death of Michael Brown at the hands of a white policeman.
The user collects two infographics with topics on murder rate across races, and the SWAT deployment rate of different races.
The online interactive visualization CommentIQ looks deeper into the comments of the article from The New York Times.

\begin{figure}[htb]
  \centering
  \frame{\resizebox{\columnwidth}{!}{\includegraphics{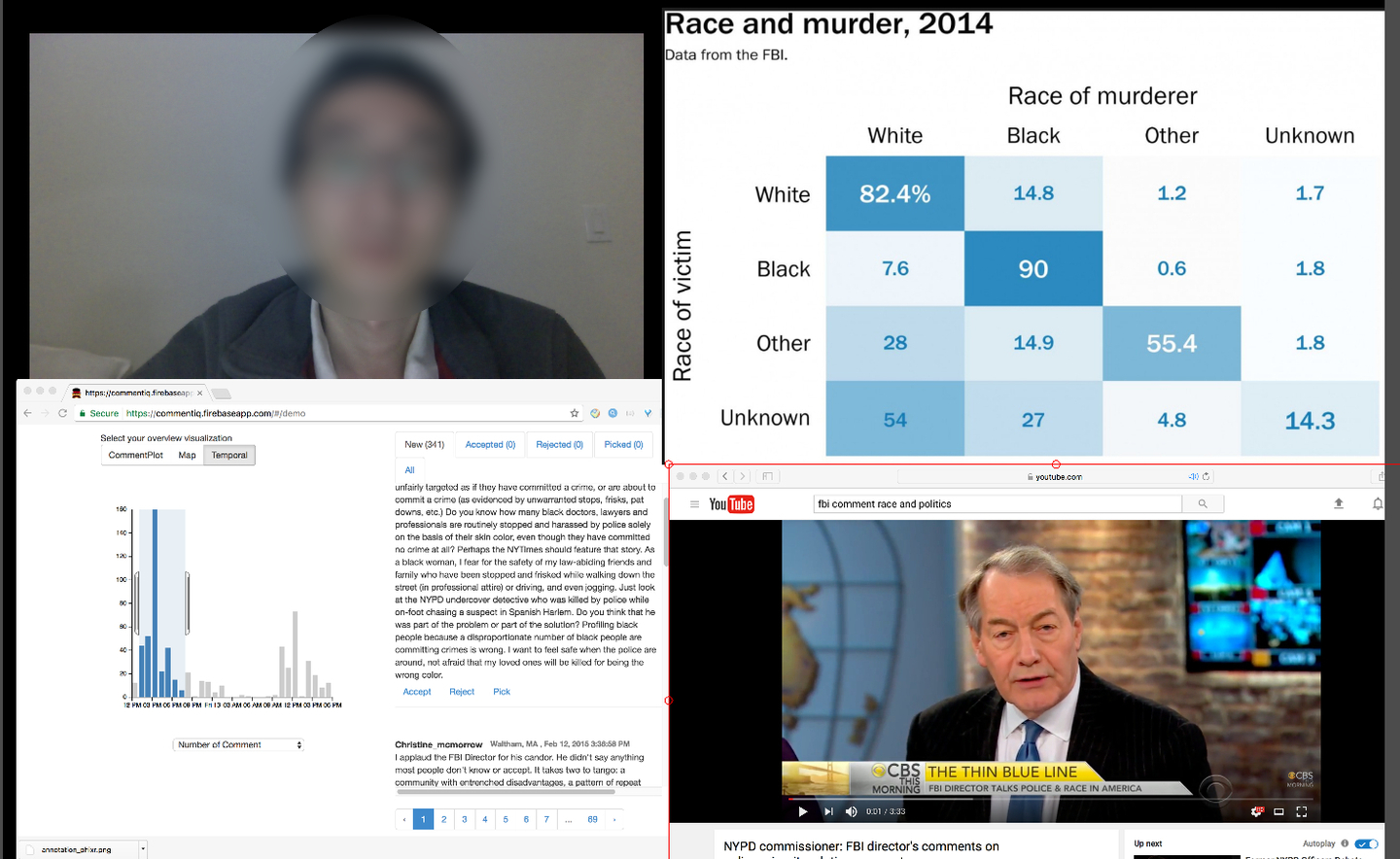}}}
  \caption{DataTV being used to record a streaming data video on race, murder rate, and SWAT activity.}
  \label{fig:commentIQ_info}
\end{figure}

This data video leverages the advantages of each of the three types of media source: video, infographics, and visualization.
First, the narrator uses a YouTube video of the Ferguson incident as the introduction.
This video shows the confrontation between the protesters and the police, providing a suitable framing to the video that emphasizes the direness of the situation.
The two infographics (Figure~\ref{fig:commentIQ_info}) give background information by showing an overview of the guns and crimes in the area.
Finally, the CommentIQ visualization allows the narrator to discover trends in how the NYT commenter community responded to the article.
In all of these cases, the narrator can scribble directly on the composited video output to highlight interesting aspects of the video, such as outliers or trends.

The live streaming functionality of the DataTV platform opens up an entirely new potential for the New York Times to provide a live complement to go with their online comment system.
Using the DataTV live stream, community moderators could aggregate and discuss comments in real time, for example when polling voter panels for political debates.

\begin{figure}[htb]
  \centering
  \frame{\resizebox{\columnwidth}{!}{\includegraphics{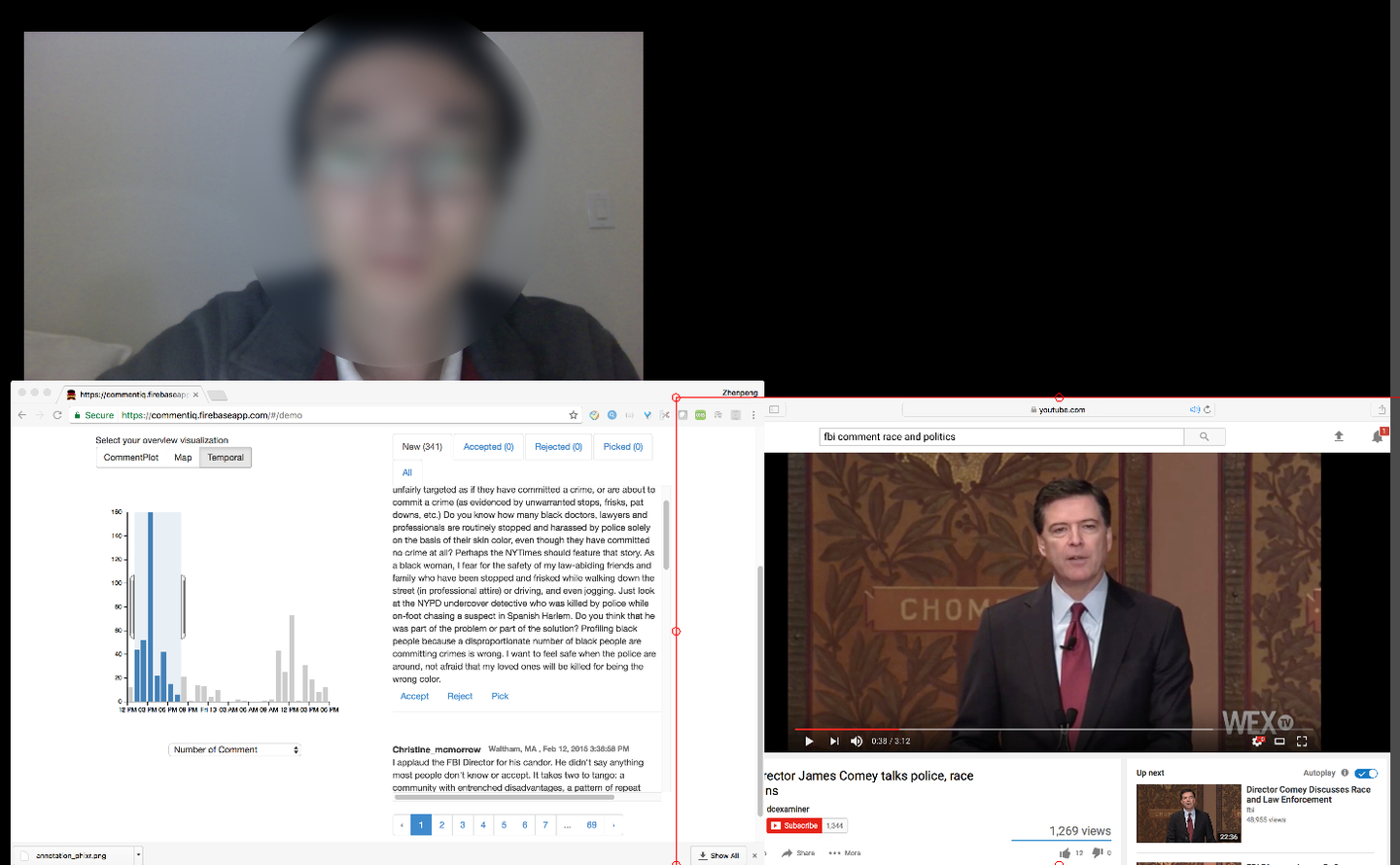}}}
  \caption{DataTV recording the CommentIQ system being used to filter comments over time.}
  \label{fig:commentIQ_temporal}
\end{figure}

%% file: 6-evaluation.tex
% 5-evaluation.tex
\section{Qualitative Evaluation}

In order to understand how experts and practitioners from the field of information science and HCI would use DataTV, and to identify potential advantages and challenges in using DataTV, we conducted a usability study.
Because there is no comparable tool for creating live-streamed data videos in real time, we opted to not perform a comparative study, but instead to focus on the affordances and capabilities of DataTV in a qualitative evaluation.

Our intent with the evaluation was to understand how DataTV can be used in different scenarios.
We thus picked two separate data representations, and designed a set of tasks for each data representation such that the tasks were of similar complexity across the two scenarios.
Each participant was assigned one data representation, with the associated set of tasks.
During the process we use the basic expert review method proposed by Tory and M\"oller~\cite{Tory2005}, which includes experts evaluating a tool using pre-defined heuristics.
The purpose of the study was to test the usability under the context of the works of the experts. 

\subsection{Participants}

We engaged two volunteer participants---Expert 1 and 2---to evaluate our system. 
The participants were Ph.D.\ students in the field of HCI or information visualization with at least three years of experience.
Both were male.

\subsection{Apparatus}

We conducted the experiment on a standard laptop computer equipped with a 15-inch LCD screen (resolution 1280$\times$800), a standard keyboard, and a three-button mouse.
The built-in camera was used for recording the user's speech with voice.

\subsection{Tasks}

Each participant's task was to create data videos using DataTV and the data visualization randomly assigned to them.
Participants were required to use at least one interactive visualization in their video.
They were allowed to pick any other appropriate media sources on the web to support their stories.
The tasks were devised so that the experts would need to explore the visualizations, select media sources, and sift through information to create a narrative.
The data stories we asked them to create were inspired by Section~\ref{sec:examples}:

\begin{itemize}
\item\textit{\textbf{Obama Budget:}} A visualization illustrating the components of government budget in the year 2013. 
\item\textit{\textbf{U.S.\ Census:}} Demographics for the state of Florida. 
\end{itemize}

A set of tasks was prepared for each scenario, requiring the participants to explore the data visualizations in detail and to look for supporting information online, before finally creating a data video to answer the question.
A sample set of tasks for one of the scenarios (Obama budget) is given below:

\begin{itemize}
\item Explain roughly how the total (\$3.7 trillion) were allocated (using a 30-second video);
\item Explain the main types of spending (using a 30-second video);
\item Explain how the spending has changed since the last budget (using a 40-second video); and
\item Describe how much of the budget was allocated to social security (using a 60-second video).
\end{itemize}

\subsection{Procedure}

Participants were shown a demonstration of DataTV and given as much time as they needed to explore and familiarize themselves with the system.
We then gave them a sheet of tasks and asked them to create a data video in response to each task.
The resulting data video would ideally make use of multiple media types, interactive visualization, and innovative storytelling techniques.
During the creation process, which was capped at 60 minutes, the experts were allowed to ask questions about the interface.
We followed a think-aloud protocol with the participants, and recorded their behavior via video and written observations.

For the purpose of this study, the participant-generated stories were recorded and saved (instead of streamed dynamically).
The final participant-generated videos were limited to a duration of one minute.
During the process, the participants used the DataTV tool to design, sketch, record, and produce their videos.
After completing their tasks, we followed up with an interview where participants explained their process and provided feedback on the system.

\subsection{Results}

We collected and analyzed both the products as well as the observations and interview feedback from each expert review session.

\subsubsection{Products}

\begin{figure}[htb]
  \centering
  \frame{\resizebox{\columnwidth}{!}{\includegraphics{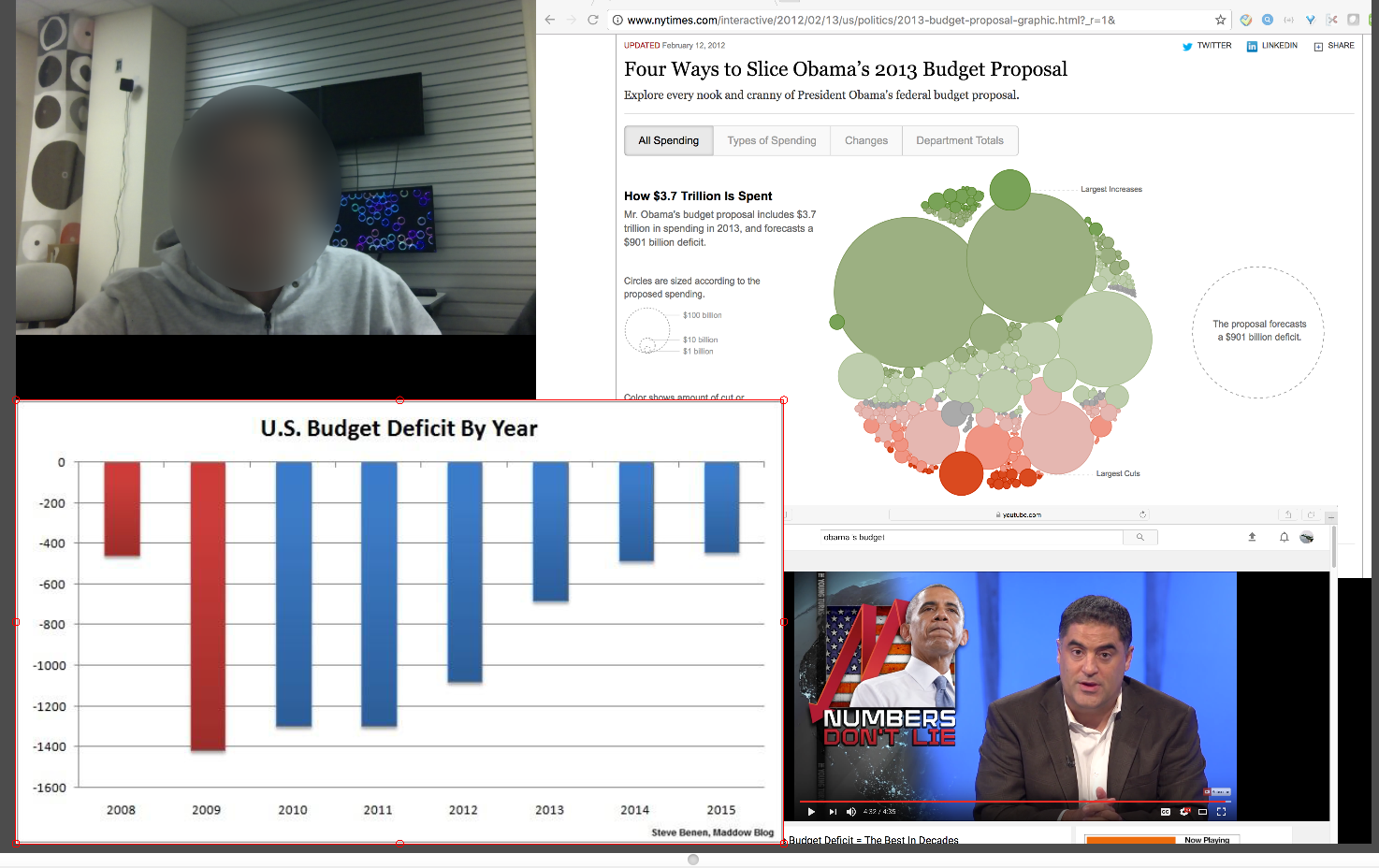}}}
  \caption{Expert 1 reviewed the U.S.\ budget trend in 2013.
  The expert presented their insights using DataTV.}
  \label{fig:budget-demo}
\end{figure}

\begin{figure}[htb]
  \centering
  \frame{\resizebox{\columnwidth}{!}{\includegraphics{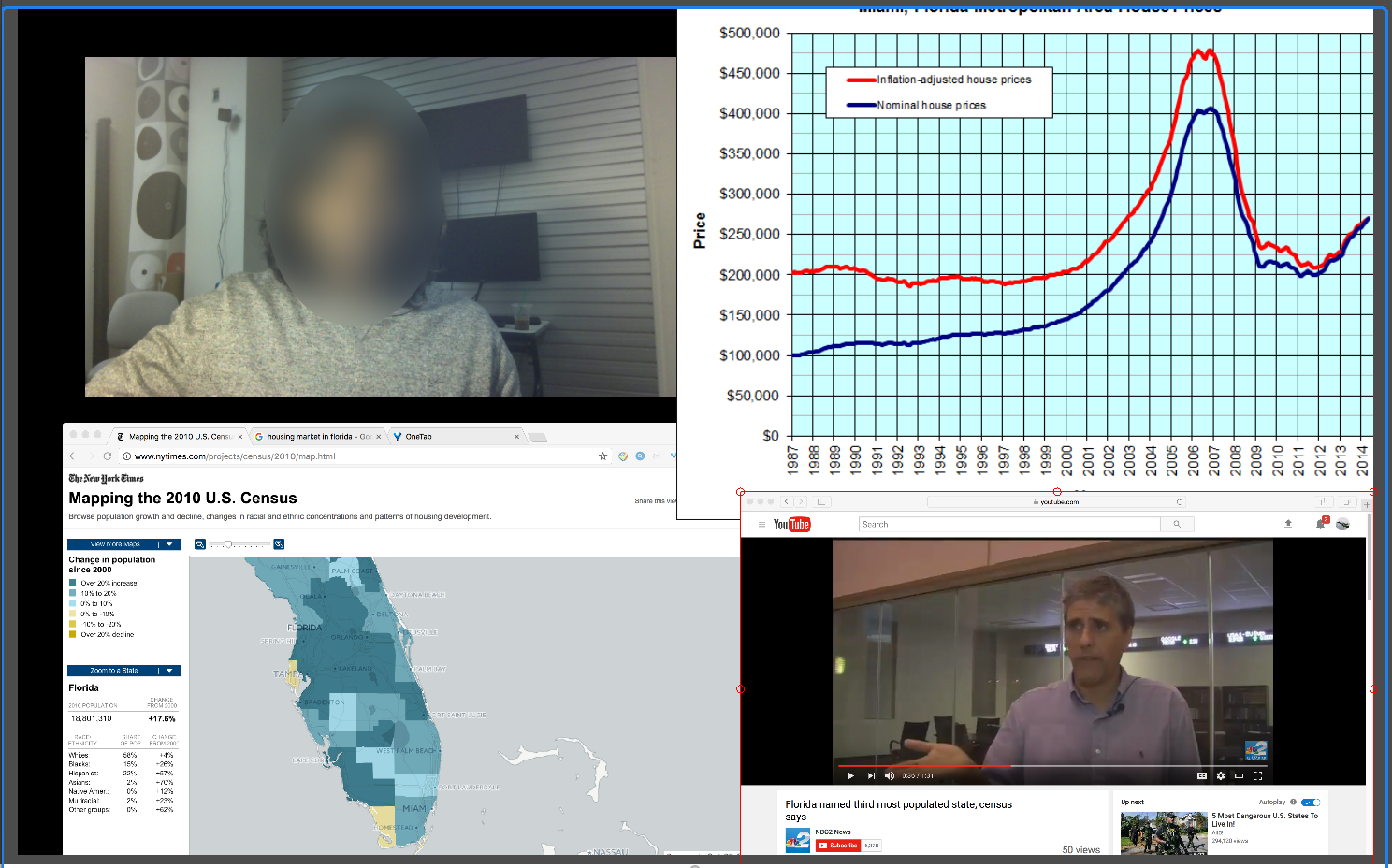}}}
  \caption{Expert 2 reviewed the population change of Florida.
  Insights on housing and population gain were presented using DataTV.}
  \label{fig:population-demo}
\end{figure}

Each expert made several data videos, all less than one minute long.
Representative data videos are attached as supplemental materials.
We also captured screenshots of the ending exact workspace for each expert (Figure~\ref{fig:budget-demo}, Figure~\ref{fig:population-demo}).
All in all, the resulting data videos were all of good quality and suggest that the DataTV platform was instrumental in the process.

\subsubsection{Observations}

Overall, participants used the tool with little training.
Both experts familiarized themselves with the basic operation by first making a few trial videos that were recorded and saved locally.
Once satisfied with the basic mental model, they spent a considerable time (20-25 minutes) finding source materials, selecting visualizations, and writing notes.
They then spent an additional 10 minutes to create their scenes, including arranging the layout and size of the different media sources. 
The actual recording of each video was surprisingly quick; given the preparation, the experts were both able to record their videos in a single take and with few mistakes.

From our observations, it appeared as if the experts did not need much prompting to get familiar with the DataTV interface. 
The experts seemed to think the interface behaved in a logical and predictable fashion, and they quickly became proficient with very little instruction and within 10 minutes of training.
Most importantly, observations and think-aloud remarks seem to indicate that the experts rapidly internalized the DataTV controls and were able to focus on the craft of data-driven storytelling.
This was indicated by their utterances increasingly dealing with how to best organize and present insights rather than minutae of the interface.

\subsubsection{Interview Feedback}

During the structured interview after completing the tasks, the experts were asked to give feedback on the system, including advantages and disadvantages of DataTV over existing approaches. 
We summarize their feedback below:

\begin{itemize}
\item\textbf{Positive:} Expert 1 thought that the DataTV interface was simple and straightforward, with few opportunities to make mistakes.
Expert 2 remarked that compared to using multiple software platforms, our system makes the video creation process seamless as it requires less operations and vital tools such as annotation, compositing, and transformations are very accessible. 

Expert 2 also thought that the use of our system was surprisingly easy, particularly the streamlined workflow where very little preparation is necessary.
The expert noted that the composite video output was helpful so as to always know what is being streamed and recorded, striking a good balance between real-time control and accuracy.

\item\textbf{Negative:} Both experts remarked on the lack of video editing capabilities. 
We responded with the fact that such editing functionality would have precluded real-time streaming of the tool, and they both remarked that the compromise was acceptable.
The first expert complained that the DataTV interface should provide better control over media sources in the workspace.

\end{itemize}

%% file: 7-discussion.tex
% 7-discussion.tex
\section{Discussion}

Our informal evaluation indicated that DataTV facilitated live data-driven storytelling.
However, the main contribution of this paper is the method of live-streaming data videos, whereas our implementation is merely a prototype to show the validity of the concept.

It is important to note that all of the functionality of the DataTV tool can be replicated in a combination of desktop recording tools---such as VLC---and video editing tools---such as Adobe Premiere Pro---with sufficient time and effort.
While DataTV makes constructing data videos easy with its integration of media source picking functionality, media label editing, and video recording, each of the DataTV videos showcased in this paper can be built using other tools.
However, the argument for the DataTV platform and related software is two-fold: (1) a single unified platform is needed to allow for live streaming and rapid production, and (2) the integration of all of these data-driven storytelling features in a single tool enables the analyst to think about data videos more in terms of storytelling rather than low-level software, mechanics, and features.
Results from our qualitative evaluation support these two arguments.

Having said all this, we believe our work surfaces several new issues that were not considered in the past.
For example, while Amini et al.\ already suggested the data videos concept in 2015~\cite{Amini2015}, their work still results in a static and prerecorded video, not a live-streamed one. 
One of the benefits may be that it is easier to quickly produce a data video using DataTV than painstakingly using a suite of tools such as screen recorders, video editors, and audio production tools.
However, DataClips~\cite{Amini2017}, presented in 2017, does provide functionality for quickly assembling several clips using predefined visualizations.
On the other hand, a live data video can be responsive to an audience, for example in responding to questions or requests for more information.
In this way, DataTV is much more of an interactive presentation tool than typical data video production tools (such as DataClips).
This is reinforced by the emphasis on live video in DataTV, whereas the narrator is typically disembodied in most existing data videos. 
We think this suggests that live data videos as those supported in our work is a unique data-driven storytelling medium in its own right.
It is also the reason that we found no easy baseline for a comparative evaluation.
We leave comparisons to live presentation software, such as Microsoft PowerPoint, to future work.

Our work in this paper has several limitations.
First of all, much of our argumentation of using data video for storytelling is based on two assumptions: that the audience has (a) enough knowledge for understanding the data video, and (b) a favorable opinion about video storytelling.
Without enough knowledge, much of the benefit of an established common ground in the visual language of data video is lost.
Furthermore, given the sometimes the higher requirement of environment---playing video might be inappropriate in some communication situations---data video might not be the perfect choice for information distribution under situation that noise level is sensitive or displaying device is not well equipped. 
For example, it can be argued that a DataTV may not be the best vehicle for presentations in very noisy settings, such as a couple people discussing a topic in a train station, where static material might be more suitable.
It should also be noted that the content of DataTV, often including personal webcam video, can be inappropriate for public broadcasting. 

Finally, the utility of live-streaming data videos as a concept can be questioned.
It is certainly true that we do not foresee ``Let's Analyze'' videos to dethrone the ``Let's Play'' category on Twitch or YouTube anytime soon.
However, the power of the internet as both a medium as well as an audience should not be underestimated.
There is already a small but growing group of Twitch communities devoted to non-gaming, such as painting, gardening, and programming.
The step is not too far from such topics to data analysis. 
Besides, even if live data videos never become popular, many of the real-time authoring techniques pioneered in DataTV will be invaluable for creating normal, non-streaming data videos, going beyond what even tools such as DataClips~\cite{Amini2017} can do.

%% file: 8-conclusion.tex
\section{Conclusion and Future Work}

We have presented a new approach to streaming data exploration in real-time.
We have also implemented a prototype desktop application (DataTV) that includes a source picker for managing media sources, a live navigation and
annotation toolbar, a composite space for aggregating components, and controls for controlling video streaming and playback. 
Results from an expert review show that DataTV encourages rapid visual storytelling and helps storytellers to consider narrative rather than video production mechanics.

Our future work will continue exploring other ways to improve narrative data videos based on DataTV.
We are also interested in exploring other storytelling media such as sketches, cartoons, technical drawings, infographics, and oral storytelling.